\documentclass[%
 amsmath,
amssymb,
pra,
reprint,
superscriptaddress
]{revtex4-1}

\usepackage{graphicx}
\usepackage{dcolumn}
\usepackage{bm}
\usepackage{braket}
\usepackage{mathptmx}

\usepackage{xcolor}
\usepackage{ulem}

\begin{document}
             
\title{Temporal-mode continuous-variable 3-dimensional cluster state \\
for topologically-protected measurement-based quantum computation}
\author{Kosuke Fukui} 
\author{Warit Asavanant}
\author{Akira Furusawa}
\affiliation{%
Department of Applied Physics, School of Engineering, The University of Tokyo,\\
7-3-1 Hongo, Bunkyo-ku, Tokyo 113-8656, Japan}                               

\begin{abstract}
Measurement-based quantum computation with continuous variables in an optical setup shows the great promise towards implementation of large-scale quantum computation, where the time-domain multiplexing approach enables us to generate the large-scale cluster state used to perform measurement-based quantum computation.
To make effective use of the advantage of the time-domain multiplexing approach, in this paper, we propose the method to generate the large-scale 3-dimensional cluster state which is a platform for topologically protected measurement-based quantum computation. Our method combines a time-domain multiplexing approach with a divide-and-conquer approach, and has the two advantages for implementing large-scale quantum computation.
First, the squeezing level for verification of the entanglement of the 3-dimensional cluster states is experimentally feasible.
The second advantage is the robustness against analog errors derived from the finite squeezing of continuous variables during topologically-protected measurement-based quantum computation.
Therefore, our method is a promising approach to implement large-scale quantum computation with continuous variables.
\end{abstract}

\maketitle
\section{Introduction}\label{Intro}
Quantum computation has a great deal of potential to efficiently solve some hard problems for conventional computers~\cite{Shor,Grov}.
To realize large-scale quantum computation, measurement-based quantum computation (MBQC) is one of the most promising quantum computation model, where universal quantum computation can be implemented with only adaptive and local single-qubit measurements on a large-scale cluster state \cite{Brie,Rau3}.
Among the candidates for quantum states, continuous variables in an optical system have shown a great potential for the generation of large-scale cluster states.
In fact, the generation of large-scale 1- and 2-dimensional cluster states has been reported in Refs.~\cite{Yoko,Yoshi} and Refs.~\cite{Warit,Lars}, respectively, where universal MBQC with continuous variables is performed on the 2-dimensional cluster state \cite{Meni6}.
More recently, arbitrary single-mode Gaussian operations over one hundred steps have been demonstrated in Ref.~\cite{Warit2}.
The ability to generate a large-scale entanglement generation comes from the fact that squeezed vacuum states can be entangled with only beam-splitter coupling through the time-domain multiplexing approach which allows us to miniaturize optical circuits~\cite{Meni5} and generate unlimited resource regardless of the coherence time of the system.
In addition, a frequency-encoded continuous variable in an optical setup is also a promising platform \cite{Meni8,Phys,Chen,Ros}, where the entangled state composed of more than 60 qumodes has been observed \cite{Phys}. 

Regarding fault-tolerant MBQC, the quantum error correction using the GKP qubit \cite{GKP} will be performed on the large-scale cluster state \cite{Meni}. In the quantum error correction with the GKP qubit, a standard quantum error correcting code such as the Steane's 7-qubit code \cite{Stea} is performed on the 2-dimensional cluster state. Alternatively, topologically protected MBQC has attracted much attention due to its high-noise threshold in implementing fault-tolerant MBQC \cite{Rau1,Rau2}. In topologically protected MBQC, a surface code \cite{Kitaev} is performed on a Raussendorf-Harrington-Goyal lattice, which is referred to as the topological cluster state in this work. In the continuous-variable system, there are many studies on the cluster state for a surface code with continuous-variables \cite{Zhang08,Milne12,Morimae13,Demarie14,Menicucci18}.
However, to the best of our knowledge, the specific method for generating the large-scale topological cluster state with continuous variables has not been studied so far.

In this paper, we propose a novel method to generate the large-scale topological cluster state, where a time-domain multiplexing approach is combined with a divide-and-conquer approach. Our method has the two advantages for implementing large-scale quantum computation. First, our method shows experimentally feasible squeezing level for verifying the entanglement of the topological cluster state, since the required squeezing level for the topological cluster state is almost the same level with the 2-dimensional cluster state generated by using only a time-domain method.
Second, our method provides the noise tolerance against analog errors derived from the finite squeezing during MBQC, where the noise propagation can be reduced thanks to the feature of the generated topological cluster state.

The rest of the paper is organized as follows. In Sec. \ref{Sec2}, we briefly review the background knowledge regarding the cluster states and measurement-based computation with continuous variables.
In Sec. \ref{Sec3}, we propose the method to generate the topological cluster state.
In Sec. \ref{Sec4}, we analyze the condition of the entanglement of the generated topological cluster state and the error propagation in topologically protected MBQC, showing two advantages of our method for implementing large-scale quantum computation. Sec.\ref{Sec5} is devoted to discussion and conclusion.
\begin{figure*}[t]
\begin{center}
 \includegraphics[angle=0, width=2.0\columnwidth]{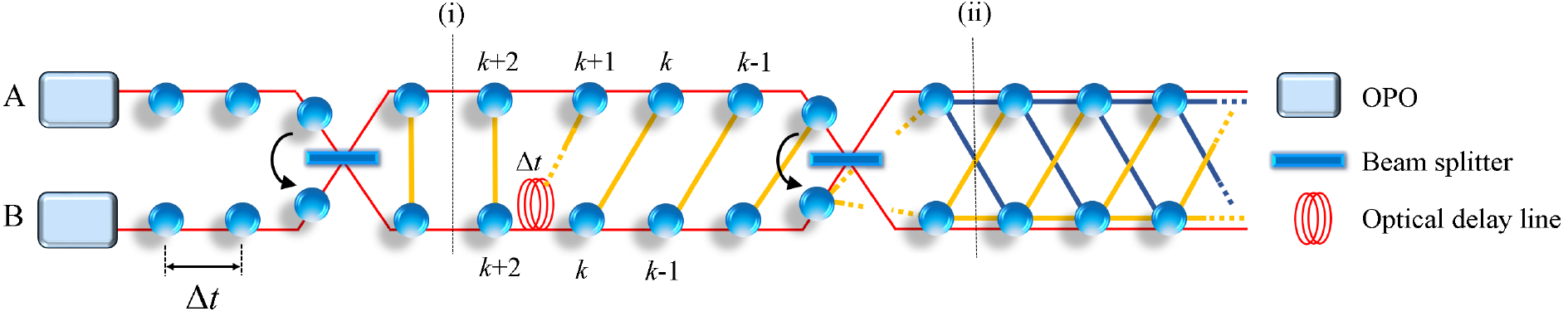} 
\caption{The generation of the 1-dimensional cluster state in an optical setup by using a time-domain multiplexing approach. Each colored circle represents a qumode, while each link between qumodes represents quantum entanglement. 
The color of the link denotes the sign of the edge-weight factor for a weighted continuous-variable cluster state; i.e., the blue and yellow edges represent + and - signs, respectively. The arrow represents the phase relationship of the unitary matrix for the beam-splitter coupling.
(i) The generation of the two-mode entangled states by using a beam-splitter coupling between a sequence of modes ${\rm A}_{k}$ and ${\rm B}_{k}$. (ii) The generation of the 1-dimensional cluster state referred to as the extended EPR state by using a beam-splitter coupling between the modes ${\rm A}_{k}$ and ${\rm B}_{k-1}$ after the time delay for mode B. The time delay, $\Delta t$, is implemented by an optical delay line.}
\label{fig1}
\end{center}
\end{figure*}
\section{Cluster state with continuous variables}\label{Sec2}
In this section, we describe the background regarding the generation of the cluster state with continuous variables by using a time-domain multiplexing. Specifically, we see as an example 1-dimensional cluster state \cite{Yoko,Yoshi} and the nullifiers \cite{Gu, Meni7} to characterize the generated cluster state.

In a continuous-variable system, position and momentum operators are defined as
\begin{equation}
\hat{q}=\frac{1}{\sqrt{2}}( \hat{a} + \hat{a}^{\dag}), \hspace{10pt}
\hat{p}=\frac{1}{i\sqrt{2}}( \hat{a} - \hat{a}^{\dag}),
\end{equation}
where $\hat{a}$ and $\hat{a}^{\dag}$ are annihilation and creation operators, and commutator relations are [$\hat{a},\hat{a}^{\dag}$] =1 and [$\hat{q},\hat{p}$]=$i$ with $\hbar=1$.
To describe the cluster states with continuous variables in an optical setup, we focus on the 1-dimensional cluster generation demonstrated in Refs. \cite{Yoko,Yoshi}, which is referred to as the extended EPR state.
Fig. \ref{fig1} shows that a large-scale 1-dimensional cluster state generated by using the time-domain multiplexing approach is composed of the squeezed vacuum states.
Temporally localized wave packets from two  optical parametric oscillators (OPOs) are used as qumodes for MBQC. In Fig. \ref{fig1}, each colored circle and link represents a qumode and the quantum entanglement, respectively, and the color of the link describes the sign of the edge-weight factor for a weighted continuous-variable cluster state \cite{Meni7,Meni5, Yoko}.
Firstly, the two-mode entangled states are generated by a beam-splitter coupling between a sequence of temporal qumodes A and B, as shown in Fig. \ref{fig1}({\rm i}), where qumodes A and B have the position and momentum squeezing, respectively.
The $k$-th mode operators for the qumodes A and B are represented by
\begin{eqnarray}
\hat{a}^{(0)}_{{\rm A},k}=({\rm e}^{-r}\hat{q}^{(0)}_{{\rm A},k}+i{\rm e}^{r}\hat{p}^{(0)}_{{\rm A},k})/\sqrt{2},\nonumber\\
\hat{a}^{(0)}_{{\rm B},k}=({\rm e}^{r}\hat{q}^{(0)}_{{\rm B},k}+i{\rm e}^{-r}\hat{p}^{(0)}_{{\rm B},k})/\sqrt{2},
\end{eqnarray}
where $\hat{q}_{\rm A(B)}^{(0)}$ and $\hat{p}_{\rm A(B)}^{(0)}$ are position and momentum quadratures of the vacuum state with the squeezing parameters $r_{\rm A(B)}$ for the $k$-th qumode A(B) , respectively.
The 50:50 beam-splitter coupling \cite{Note1} transforms the operators for qumodes ${\rm A}_{k}$ and ${\rm B}_{k}$ as
\begin{eqnarray}
\hat{U}^{\dag}_{\rm BS}\left( 
\begin{array}{c}
\hat{a}^{(0)}_{{\rm A},k}\nonumber \\
\hat{a}^{(0)}_{{\rm B},k}
\end{array}
\right)
\hat{U}_{\rm BS}&=&\frac{1}{\sqrt{2}}
\left( 
\begin{array}{cc}
1&-1\\
1&1
\end{array}
\right) \left( 
\begin{array}{c}
\hat{a}^{(0)}_{{\rm A},k}\\
\hat{a}^{(0)}_{{\rm B},k}
\end{array}
\right)\\
&=& \left( 
\begin{array}{c}
\hat{a}^{ ({\rm i})}_{{\rm A},k}\\
\hat{a}^{ ({\rm i})}_{{\rm B},k}
\end{array}
\right).
\end{eqnarray}
Secondly, the time delay for the qumode B, $\Delta t$, is implemented with an optical delay line whose length is equal to the time interval between adjacent qumodes. After the time delay, a beam-splitter coupling is performed between qumodes A and B, and transforms the operators for modes ${\rm A}_{k}$ and ${\rm B}_{k-1}$ as
\begin{eqnarray}
\hat{U}^{\dag}_{\rm BS}\left( 
\begin{array}{c}
\hat{a}^{ ({\rm i})}_{{\rm A},k}\\
\hat{a}^{ ({\rm i})}_{{\rm B},k-1}
\end{array}
\right)
\hat{U}_{\rm BS}=\frac{1}{\sqrt{2}}
\left( 
\begin{array}{cc}
1& -1\\
1&1
\end{array}
\right) \left( 
\begin{array}{c}
\hat{a}^{ ({\rm i})}_{{\rm A},k}\\
\hat{a}^{ ({\rm i})}_{{\rm B},k-1}
\end{array}
\right) \nonumber \\
=\frac{1}{2}\left( 
\begin{array}{c}
\hat{a}^{(0)}_{{\rm A},k}-\hat{a}^{(0)}_{{\rm B},k}-\hat{a}^{(0)}_{{\rm A},k-1}-\hat{a}^{(0)}_{{\rm B},k-1}\\
\hat{a}^{(0)}_{{\rm A},k}-\hat{a}^{(0)}_{{\rm B},k}+\hat{a}^{(0)}_{{\rm A},k-1}+\hat{a}^{(0)}_{{\rm B},k-1}
\end{array}
\right)=\left( 
\begin{array}{c}
\hat{a}^{({\rm ii})}_{{\rm A},k}\\
\hat{a}^{({\rm ii})}_{{\rm B},k}
\end{array}
\right).
\end{eqnarray}
After the second beam-splitter coupling, we finally obtain the 1-dimensional cluster state, as shown in Fig. \ref{fig1}({\rm ii}).

To characterize the generated 1-dimensional cluster state, we introduce the nullifiers. The nullifier corresponds to the stabilizer for cluster states with discrete variables in the case of the infinite squeezing, and is used to verify the generated cluster state.
The nullifiers of the qumode $k$ for the generated 1-dimensional cluster state in the $q$ and $p$ operators, $\hat{\delta}_{k}^{q}$ and  $\hat{\delta}_{k}^{p}$, are obtained as
\begin{eqnarray}
\hat{\delta}_{k}^{q}&=&\hat{q}^{({\rm ii})}_{{\rm A},k}+\hat{q}^{({\rm ii})}_{{\rm B},k}-\hat{q}^{({\rm ii})}_{{\rm A},k+1}+\hat{q}^{({\rm ii})}_{{\rm B},k+1},\\
\hat{\delta}_{k}^{p}&=&-\hat{p}^{({\rm ii})}_{{\rm A},k}-\hat{p}^{({\rm ii})}_{{\rm B},k}-\hat{p}^{({\rm ii})}_{{\rm A},k+1}+\hat{p}^{({\rm ii})}_{{\rm B},k+1},
\end{eqnarray}
respectively \cite{Yoko,Yoshi}. In Eqs. (5) and (6), note that the label $k$ for the qumodes B in Fig. \ref{fig1} is relabeled to $k+1$ due to the nullifier formalism (see Appendix A for details on the calculation of nullifiers for the generated 1-dimensional cluster state).
Using Eqs. (2)-(4), 
we obtain the relations as
\begin{eqnarray}
\hat{\delta}_{k}^{q}=2{\rm e}^{-r_{{\rm A}}}\hat{q}_{{\rm A},k}^{(0)},\\
\hat{\delta}_{k}^{p}=2{\rm e}^{-r_{\rm B}}\hat{p}_{{\rm B},k}^{(0)}.
\end{eqnarray}
In the case of the ideal 1-dimensional cluster state, i.e, the squeezed vacuum state has an infinite squeezing, the nullifiers for the 1-dimensional cluster state $\ket{1{\rm D}}$ become zero as
\begin{equation}
\hat{\delta}_{k}^{q}\ket{1{\rm D}}=0, \hspace{5pt} 
\hat{\delta}_{k}^{p}\ket{1{\rm D}}=0.
\end{equation}
Thus, nullifiers for the cluster state with the infinite squeezing correspond to the stabilizer.
In the case of the finite squeezing, we can verify the generation of the 1-dimensional cluster state by calculating the inseparable condition for the variance as
\begin{eqnarray}
\langle(\hat{\delta}_{k}^{q})^2\rangle={\rm e}^{-2r_{A}}\langle(\hat{q}_{{\rm A},k}^{(0)})^2\rangle<\frac{1}{2},\\
\langle(\hat{\delta}_{k}^{p})^2\rangle={\rm e}^{-2r_{B}}\langle(\hat{p}_{{\rm B},k}^{(0)})^2\rangle<\frac{1}{2},
\end{eqnarray}
where $\langle \hat{O} \rangle$ denotes the expectation value of the operator $\hat{O}$, and the variance for the vacuum states, $\langle(\hat{q}_{{\rm A},k}^{(0)})^2\rangle$ and $\langle(\hat{p}_{{\rm B},k}^{(0)})^2\rangle$, are equal to 1/2.
This condition to verify the entanglement generation is called as van-Loock-Furusawa criterion \cite{Loock1}.
From this criterion, the squeezing level required for the 1-dimensional cluster state is -3.0 dB squeezing of each nullifier, where a squeezing level is equal to 10${\rm log}_{10}{\rm e}^{-2r}$.

 \section{Generation of the topological cluster state}\label{Sec3}
\begin{figure*}[t]
\begin{center}
 \includegraphics[angle=0, width=2.1\columnwidth]{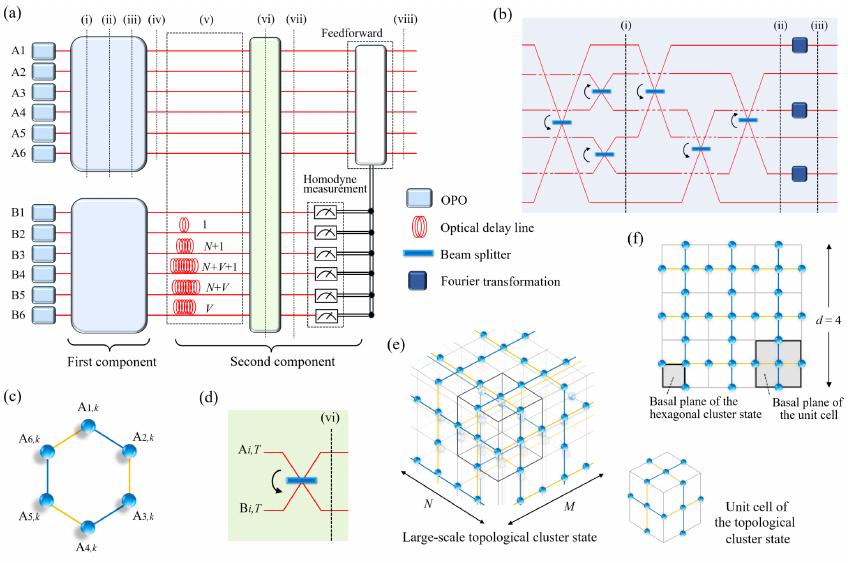} 
\caption{A schematic representation of the process of our method to generate the large-scale topological cluster state.
(a) Generation of the topological cluster state from two components. (i)-(iv) The generation of the small-scale cluster states called as the hexagonal cluster state in the first component. (v)-(viii) The generation of the large-scale topological cluster state from hexagonal cluster states using a time-domain multiplexing method, where the size of the basal plane for the topological cluster state is $V=N\times M$.
Time delays are implemented on qumodes ${\rm B}_{2,k}$, ${\rm B}_{3,k}$, ${\rm B}_{4,k}$, ${\rm B}_{5,k}$, and ${\rm B}_{6,k}$ by $1$, $N+1$, $N+V+1$, $N+V$, and $V$, respectively, assuming that the $k$-th qumode with a time delay $V\times \Delta t$ is coupled with the qumode in the $(k+V)$-th hexagonal cluster state A.
(b) Experimental setup for the first component, generating the hexagonal cluster state. Each of the generator of hexagonal cluster A and B in the first component consists of 6 optical parametric oscillators (OPOs) and 6 beam-splitters. (i) The generation of two-mode entangled states. (ii) The generation of the multimode entangled states. (iii) Fourier transformation on three qumodes. (c) The generated hexagonal cluster state A.
(d) The beam-splitter coupling between the qumodes A and B in the second component.
(e) (left) The generated large-scale topological cluster state, where the size of the basal plane for the space-like direction is $N\times M$, and the length for the time-like direction is arbitrarily large.
The large-scale topological cluster state is generated via the quantum erasure, i.e., the measurement of qumodes B by using the homodyne measurement and the feed-forward depending on the measurement results after (a)(vii).
 (right) The unit cell of the topological cluster state. (f) Basal planes of the topological cluster state with the distance of the array $d=4$ in terms of the one horizontal slice at perpendicular to the time direction. The distance of the array for a surface code, $d$, corresponds to $M$, assuming $M=N$.
}
\label{fig2}
\end{center}
\end{figure*}
In this section, we propose the method to generate the topological cluster state whose squeezing level required for the entanglement is experimentally accessible to date.
In our method, the so-called divide-and-conquer approach \cite{Niel, Daws} is combined with the time-domain method.
We note that the purpose of using the divide-and-conquer approach in Refs. \cite{Niel, Daws} is to overcome a problem based on a photon qubit in terms of the probabilistic two-qubit gate for generating the large-scale cluster state, while our purpose is to achieve the feasible squeezing level required for verifying the deterministic entanglement of the large-scale cluster state.
Regarding the nullifiers of the generated topological cluster state, we analyze those in the next section.

Fig. \ref{fig2}(a) shows the schematic diagram for the experimental setup to generate the large-scale topological cluster state using a miniaturized optical setup.
The setup consists of two components. In the first component, the small-scale cluster states are generated without the time-domain multiplexing approach, where the small-scale cluster states are generated from two generators.
In the second component, the large-scale topological cluster state is generated by using the time-domain multiplexing approach~\cite{Meni5,Yoshi,Warit}. 
The generated topological cluster state is depicted in Fig. \ref{fig2}(e), where the basal plane for the space-like direction has $N\times M$ modes, and the length for the time-like direction is arbitrarily large. 
Fig. \ref{fig2}(f) represents a schematic diagram of the basal plane of the topological cluster state for the so-called distance of the array, $d=4$, for a surface code. The distance of the array corresponds to ${M}$, assuming that $M$ is equal to $N$.

We explain the first component to generate the small-scale cluster states referred to as the hexagonal cluster state in this work.
Fig. \ref{fig2}(b) shows a schematic picture of generation of the hexagonal cluster state A.
Each of generators of the hexagonal cluster state consists of six OPOs and six 50:50 beam splitters, where the transmittances of beam splitters are obtained from the decomposition technique for the beam splitter network \cite{Loock}.
The odd and even numbered qumodes from OPOs have the momentum and position squeezing, respectively. As with Eq. (2), mode operators for the odd and even numbered qumodes A (B) are represented as
\begin{eqnarray}
\hat{a}^{(0)}_{{\rm A(B)},2n-1, k}&=&({\rm e}^{r}\hat{q}^{(0)}_{{\rm A(B)},2n-1, k}+i{\rm e}^{-r}\hat{p}^{(0)}_{{\rm A(B)},2n-1, k})/\sqrt{2},\nonumber\\
\hat{a}^{(0)}_{{\rm A(B)},2n,k}&=&({\rm e}^{-r}\hat{q}^{(0)}_{{\rm A(B)},2n,k}+i{\rm e}^{r}\hat{p}^{(0)}_{{\rm A(B)},2n,k})/\sqrt{2},
\end{eqnarray} 
respectively, where $n$ =1,2,3.
Here we describe the transformation of annihilation and creation operators in the generator labeled with A. The hexagonal cluster state B is generated in the same way as the hexagonal cluster state A.
In Fig. \ref{fig2}(b)(i), the generation of the two-mode entangled states by a beam-splitter coupling between a sequence of modes $i$ and $j$ is described.
This beam-splitter coupling transforms the operators as
\begin{eqnarray}
\hat{U}^{\dag}_{\rm BS}\left( 
\begin{array}{c}
\hat{a}^{(0)}_{{\rm A}{,i,k}}\\
\hat{a}^{(0)}_{{\rm A}{,j,k}}
\end{array}
\right)
\hat{U}_{\rm BS}&=&\frac{1}{\sqrt{2}}
\left( 
\begin{array}{cc}
1& -1\\
1&1
\end{array}
\right) \left( 
\begin{array}{c}
\hat{a}^{(0)}_{{\rm A}{,i,k}}\\
\hat{a}^{(0)}_{{\rm A}{,j,k}}
\end{array}
\right)  \nonumber  \\ 
&=& \left( 
\begin{array}{c}
\hat{a}^{({\rm i})}_{{\rm A}{,i,k}}\\
\hat{a}^{({\rm i})}_{{\rm A}{,j,k}}
\end{array}
\right),
\end{eqnarray}
where the sets of indices $(i,j)$ are (1,6), (5,4), and (3,2).
In Fig. \ref{fig2}(b)(ii), the multimode entangled state are generated by a beam-splitter coupling between a sequence of modes.
After this beam-splitter coupling, the operators become
\begin{eqnarray}
\hat{U}^{\dag}_{\rm BS}\left( 
\begin{array}{c}
\hat{a}^{({\rm i})}_{{\rm A}{,i,k}}\\
\hat{a}^{({\rm i})}_{{\rm A}{,j,k}}
\end{array}
\right)
\hat{U}_{\rm BS}&=&\frac{1}{\sqrt{2}}
\left( 
\begin{array}{cc}
1& -1\\
1&1
\end{array}
\right) \left( 
\begin{array}{c}
\hat{a}^{({\rm i})}_{{\rm A}{,i,k}}\\
\hat{a}^{({\rm i})}_{{\rm A}{,j,k}}
\end{array}
\right)   \nonumber   \\
&=& \left( 
\begin{array}{c}
\hat{a}^{({\rm ii})}_{{\rm A}{,i,k}}\\
\hat{a}^{({\rm ii})}_{{\rm A}{,j,k}}
\end{array}
\right),
\end{eqnarray}
where the sets of indices $(i,j)$ are (1,4), (5,2), and (3,6).
After the Fourier transformation on modes $i$ $(i= 1, 3, 5)$ described in Fig. \ref{fig2}(b), the hexagonal cluster state is generated, as shown in Fig. \ref{fig2}(c).
The operators for the hexagonal cluster state become
\begin{eqnarray}
\hat{U}^{\dag}_{\rm F}\left( 
\begin{array}{c}
\hat{a}^{({\rm ii})}_{{\rm A}{,i,k}}\\
\hat{a}^{({\rm ii}) \dag}_{{\rm A}{,i,k}}
\end{array}
\right)
\hat{U}_{\rm F}&=&
\left( 
\begin{array}{cc}
i& 0\\
0&-i
\end{array}
\right) \left( 
\begin{array}{c}
\hat{a}^{({\rm ii}) }_{{\rm A}{,i,k}}\\
\hat{a}^{({\rm ii}) \dag}_{{\rm A}{,i,k}}
\end{array}
\right)  \nonumber \\
&=&\left( 
\begin{array}{c}
\hat{a}^{({\rm iii}) }_{{\rm A}{,i,k}}\\
\hat{a}^{({\rm iii}) \dag }_{{\rm A}{,i,k}}
\end{array}
\right),
\end{eqnarray}
whereas operators for qumodes $i$ $(i=2,4,6)$ are $\hat{a}^{({\rm ii}) }_{{\rm A}{,i,k}}=\hat{a}^{({\rm iii}) }_{{\rm A}{,i,k}}$.
In the same way as the generation of the hexagonal cluster state A in the first component, the hexagonal cluster state B is obtained at the same time with the same configuration of optical elements for the hexagonal cluster state A.

\begin{figure*}[t]
\begin{center}
 \includegraphics[angle=0, width=2.1\columnwidth]{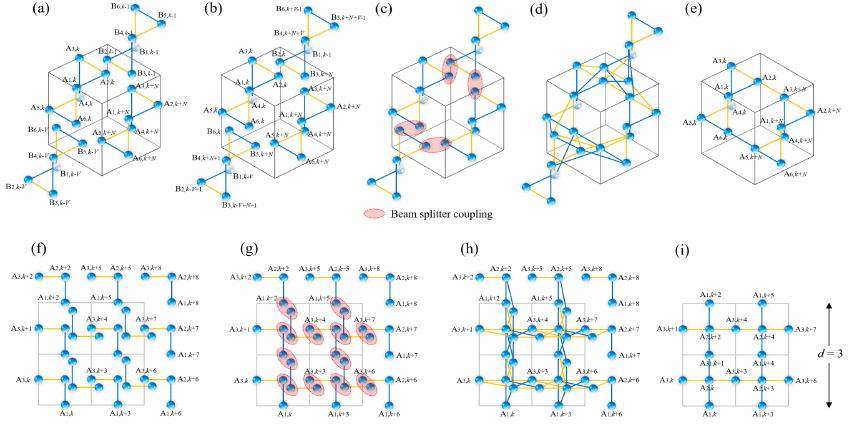} 
\caption{Introduction of generating the large-scale topological cluster state.
(a) The hexagonal cluster state A and B after the first component before a time delay. (b) The hexagonal cluster states A and B after time delays via optical delay lines. (c) Beam-splitter coupling between qumodes A and B with the same temporal mode index. (d) Entangled state consisting of hexagonal cluster states A and B. 
(e) The generated cluster state after the quantum erasure, i.e., the measurement of hexagonal cluster states B and the feed-forward operation depending on the measurement results. (f)-(i) The process of generating the topological cluster state with the distance of the array $d=3$ in terms of the one horizontal slice at perpendicular to the time direction. The qumodes and edges contained in the horizontal slice are shown.}
\label{fig3}
\end{center}
\end{figure*}

In the second component, the large-scale topological cluster state is generated by a beam-splitter coupling between qumodes A and B, and by the measurement of qumodes belonging to the hexagonal cluster state B.
In this component, the time-domain multiplexing approach is applied to hexagonal cluster states A and B in Fig. \ref{fig2}(a)(v)-(vii).
Each of modes composed of the hexagonal cluster A is coupled with the mode of the hexagonal cluster B after time delays as shown in Fig. \ref{fig2}(d).
After generating hexagonal cluster states A and B in Fig. \ref{fig2}(a)(iv), time delays are implemented to qumodes $B_{2,k}$, $B_{3,k}$, $B_{4,k}$, $B_{5,k}$, and $B_{6,k}$ by $1$, $N+1$, $N+1+V$, $N+V$, and $V$, respectively, whereas the qumodes in the hexagonal cluster states A do not have a time delay, as shown in Fig. \ref{fig2}(a)(v).
The time delays $N$ and $V(=N\times M)$ are determined by the desired lattice size of the topological cluster states, $N\times M$.
The optical delay lines 1, $N$, $M$ and $V$ are used to implement time delays $\Delta t$, $N \times \Delta t$, $M \times \Delta t$, and $V \times \Delta t= N \times M \times \Delta t$. In Fig. \ref{fig2}(a), unit of time delay, $\Delta t$, is omitted for brevity.
After the time delays, the qumodes of hexagonal clusters A and B are coupled by 50:50 beam splitters in Fig. \ref{fig2}(a)(vi). Fig. \ref{fig2}(d) shows a beam-splitter coupling between qumodes in the hexagonal clusters A and B with a same timing $T$. 
The following equation is a list for the pairs of two modes coupled by a beam splitter in terms of the $k$-th hexagonal cluster B as
\begin{eqnarray}
({\rm A},1,k) &\Leftrightarrow&  ({\rm B},1,k), \nonumber  \\
({\rm A},2,k+1) &\Leftrightarrow&  ({\rm B},2,k),\nonumber \\
({\rm A},3,k+N+1) &\Leftrightarrow&  ({\rm B},3,k),\nonumber \\
({\rm A},4,k+N+1+V) &\Leftrightarrow& ({\rm B},4,k),\nonumber \\  
({\rm A},5,k+N+V) &\Leftrightarrow& ({\rm B},5,k),\nonumber \\  
({\rm A},6,k+V) &\Leftrightarrow& ({\rm B},6,k), \label{6bs}
\end{eqnarray}
where $k=0,1,2,3,\dots.$.
The first row implies that the qumode 1 without a time delay in the $k$-th hexagonal cluster A is coupled with the qumode 1 without a time delay in the $k$-th hexagonal cluster B.
The second row implies that the qumode 2 without a time delay in the $(k+1)$-th hexagonal cluster A is coupled with the qumode 2 with a time delay $\Delta t$ in the $k$-th hexagonal cluster B.
For the third row, the qumode 3 without a time delay in the $(k+N+1)$-th hexagonal cluster A is coupled with the qumode 3 with a time delay $(N+1)\times \Delta t$ in the $k$-th hexagonal cluster B.
In the same way as the first, second, and third rows, other qumodes are coupled by using the beam splitter.
The beam-splitter coupling for the first row transforms the operator as
\begin{eqnarray}
\hat{U}^{\dag}_{\rm BS2}\left( 
\begin{array}{c}
\hat{a}^{({\rm iii})}_{{\rm A}_{1,k}}\\
\hat{a}^{({\rm iii})}_{{\rm B}_{1,k}}
\end{array}
\right)
\hat{U}_{\rm BS2}&=&\frac{1}{\sqrt{2}}
\left( 
\begin{array}{cc}
1& 1\\
1&-1
\end{array}
\right) \left( 
\begin{array}{c}
\hat{a}^{({\rm iii})}_{{\rm A}_{1,k}}\\
\hat{a}^{({\rm iii})}_{{\rm B}_{1,k}}
\end{array}
\right)   \nonumber   \\
&=& \left( 
\begin{array}{c}
\hat{a}^{({\rm iv})}_{{\rm A}_{1,k}}\\
\hat{a}^{({\rm iv})}_{{\rm B}_{1,k}}
\end{array}
\right),
\end{eqnarray}
where we use the unitary matrix $\hat{U}_{\rm BS2}$, different from $\hat{U}_{\rm BS}$, for the beam-splitter coupling between qumodes A and B.
Other operators are transformed in the same way as the first row described in Eq. (16).
We here note that the label $k$ is used for a timing in order to describe the time delays. For example, after the time delay $N\times \Delta t$ on the qumode 3 in the $(k-N)$-th hexagonal cluster B, the $q$ operator for the qumode 3, $\hat{q}_{{\rm B},3,k}$, becomes $\hat{q}_{{\rm B},3,k}$, and then the qumode 3 in the cluster B is coupled with the qumode 3 in the cluster A, $\hat{q}_{{\rm A},3,k}$.

After the beam-splitter coupling between the qumodes A and B, the large-scale entangled state, not the topological cluster state, is generated in Fig. \ref{fig2}(a)(vii).
To obtain the large-scale topological cluster state, the qumodes B need to be removed from the large-scale entangled state by using the so-called {\it quantum erasure} which has been demonstrated in Ref. \cite{Miwa}.
The quantum erasure is used for the decoupling of unwanted qumodes from a fixed large-scale cluster state by measuring the unwanted qumodes and performing the feed-forward operation depending on measurement results on the neighboring qumodes \cite{Gu}.
In our case, the qumodes B are measured by the homodyne measurement in the $q$ quadrature and the feed-forward operation is performed on qumodes A, as shown in Fig. \ref{fig2}(a) (viii).
Finally, we obtain the large-scale cluster state as depicted in Fig. \ref{fig2}(e).

The basal plane and the vertical axis of the topological cluster state are used for the space-like and time-like directions, respectively \cite{Rau1}. The size of the basal plane for the space-like direction is $N\times M$ within finite coherence time of the light source, and the length for the time-like direction is arbitrarily large. 
We note that during the MBQC the qumodes 4, 5, and 6 in the first $V$ hexagonal clusters A will be measured in the $q$ quadrature, since those do not couple with any other qumodes, and do not compose the topological cluster state. 

To get a more intuitive understanding of using the time-domain multiplexing method, we describe the schematic view of generating and entanglement between neighboring hexagonal cluster states A in Fig. \ref{fig3}(a)-(e), and the  the process of generating the topological cluster state with the distance of the array $d=3$ in Fig. \ref{fig3}(f)-(i).
In Fig. \ref{fig3}(a)-(e), we here focus on two hexagonal cluster states A whose time delay is $N\times \Delta t$, and see the entanglement generation between them via two hexagonal cluster states B with a time delays.
Figs. \ref{fig3}(a) and (b) show four hexagonal cluster states before and after time delays, respectively.
Then, beam-splitter coupling between qumodes A and B with the same temporal mode index is implemented, as shown in Fig. \ref{fig3}(c), where the beam-splitter coupling is depicted by dotted lines.
The entangled state is generated with four hexagonal cluster states after the beam-splitter coupling, as shown in Fig. \ref{fig3}(d).
Then, we implement the quantum erasure; namely, the qumodes B are measured in the $q$ quadrature and the feed-forward operation depending on the measurement results is implemented on qumodes A.
Removing the qumodes B through the erasing technique is needed to implement topologically protected MBQC, where the qumode A is measured in the $p$ quadrature to implement the quantum error correction with a surface code.
After the quantum erasing, the cluster state, which is a part of the topological cluster state, is generated, as shown in Fig. \ref{fig3}(e). 

In Fig. \ref{fig3}(f)-(i), we can see the process of generating the topological cluster state with the distance $d=3$ in terms of the one horizontal slice at perpendicular to the time direction. For simplicity, we describe only qumodes and edges contained in the horizontal slice. Note that the qumodes A, which are located in outside of the upper and right sides of the basal plane, are not needed to implement the topologically protected MBQC. Thus, some of the qumodes A, e.g., ${\rm A}_{2, \nu  M-1}$, ${\rm A}_{3, \nu  M-1}$, ${\rm A}_{1, M(N-1)+\upsilon}$, and ${\rm A}_{2, M (N-1) +\upsilon}$, are removed by using the quantum erasing, where $\nu=1,\dots.N$ and $\upsilon=0,1,\dots.M-1$, as shown in Fig. \ref{fig3}(i). In addition, some of the hexagonal cluster states B, which correspond to qumodes located on outside of the upper and right sides of the basal plane, do not contribute to the generation of the topological cluster state. Therefore, we would not generate them in the first component in Fig. \ref{fig2}. (a). These additional operations are easy to perform in our setup.

 \section{Analysis}\label{Sec4}
In this section, we firstly analyze the nullifiers of the qumodes composed of generated hexagonal and topological cluster states generated by the proposed method. We then describe the verification of the generated topological cluster state by using the nullifiers, and obtain the required squeezing level for the verification.
We finally show a robustness against analog errors in generated states by describing the fact that errors in the $q(p)$ quadrature, which are derived from the finite squeezing, do not propagate on the basis in the $p(q)$ quadrature between qumodes.

\subsection{Nullifier of the topological cluster state}
We firstly describe the nullifier of the generated hexagonal cluster state, which obeys the transformations described in Eqs. (13)-(15).
In the following, we see the generation of the hexagonal cluster state A.
Since the odd and even numbered qumodes from OPOs have the momentum and position squeezing, respectively, the initial nullifiers for the 6 modes in the temporal mode index $k$ are described as
\begin{equation}
\{ \hat{p}_{{\rm A},2n-1,k},\hspace{3pt}\hat{q}_{{\rm A},2n,k}\},
\end{equation}
where $n$ = 1,2,3. 
For sake of simplicity, we omit labels A and $k$ in Eq. (18) as $\{ \hat{p}_{2n-1},\hspace{3pt}\hat{q}_{2n}\}.$
The nullifiers for the entangled states after the first beam-splitter coupling become 
\begin{eqnarray}
\{\hat{p}_{2n-1}+ \hat{p}_{2n+4\hspace{1pt}{\rm mod}\hspace{1pt}6},\hspace{3pt}\hat{q}_{2n}-\hat{q}_{2n+1\hspace{1pt}{\rm mod}\hspace{1pt}6}\}.
\end{eqnarray}  
In Eq. (19), for instance, the nullifier for the qumode 1 changes from $ \hat{p}_{1}$ to $ \hat{p}_{1}+\hat{p}_{6}$ after the first beam-splitter between qumodes 1 and 6.
We then perform the second beam-splitter coupling in Fig. \ref{fig2}(b)(ii), and obtain nullifiers as
\begin{eqnarray}
\{ \hat{p}_{2n-1}-\hat{p}_{2n+1\hspace{1pt}{\rm mod}\hspace{1pt}6}+\hat{p}_{2n+2\hspace{1pt}{\rm mod}\hspace{1pt}6}+\hat{p}_{2n+4\hspace{1pt}{\rm mod}\hspace{1pt}6},\nonumber \\
\hat{q}_{2n}-\hat{q}_{2n+4\hspace{1pt}{\rm mod}\hspace{1pt}6}-\hat{q}_{2n+1\hspace{1pt}{\rm mod}\hspace{1pt}6}-\hat{q}_{2n+3\hspace{1pt}{\rm mod}\hspace{1pt}6}\}.
\end{eqnarray}  
After Fourier transformations on modes 1, 3, and 5 in Fig. \ref{fig2}(b)(iii), the nullifiers are transformed as
\begin{eqnarray}
\{- \hat{q}_{2n-1}+\hat{q}_{2n+1\hspace{1pt}{\rm mod}\hspace{1pt}6}+\hat{p}_{2n+2\hspace{1pt}{\rm mod}\hspace{1pt}6}+\hat{p}_{2n+4\hspace{1pt}{\rm mod}\hspace{1pt}6},\nonumber \\
\hat{q}_{2n}-\hat{q}_{2n+4\hspace{1pt}{\rm mod}\hspace{1pt}6}-\hat{p}_{2n+1\hspace{1pt}{\rm mod}\hspace{1pt}6}-\hat{p}_{2n+3\hspace{1pt}{\rm mod}\hspace{1pt}6}\}.
\end{eqnarray} 
By taking linear combinations, the nullifiers become
\begin{eqnarray}
\{ &\hat{p}&_{2n-1}+\hat{q}_{2n}-\hat{q}_{2n+4\hspace{1pt}{\rm mod}\hspace{1pt}6},\nonumber \\
&\hat{p}&_{2n}-\hat{q}_{2n+1\hspace{1pt}{\rm mod}\hspace{1pt}6}+\hat{q}_{2n+5\hspace{1pt}{\rm mod}\hspace{1pt}6}\},
\end{eqnarray}  
which corresponds to the nullifiers for the hexagonal cluster state described in Fig. \ref{fig2}(c).
In the same way as the hexagonal cluster A, the nullifiers for the hexagonal cluster B are obtained.

We next explain the nullifier of the generated topological cluster state, which obeys the transformations described in Eqs. (16) and (17).
As shown in Sec. III, the topological cluster state is generated from hexagonal clusters A and B by using the time-domain multiplexing approach, which leads to reduction of the requirement for an experimental setup to generate large-scale cluster states.
In the time delays described in Fig. \ref{fig2}(a)(v) and Eq. (\ref{6bs}), for instance, the nullifier for qumode B$_{1,k}$ changes from $\hat{p}_{{\rm B},1,k}+\hat{q}_{{\rm B},2,k}-\hat{q}_{{\rm B},6,k}$ to $\hat{p}_{{\rm B},1,k}+\hat{q}_{{\rm B},2,k+1}-\hat{q}_{{\rm B},6,k+V}$, since we are delaying qumodes B$_{2,k}$ and B$_{6,k}$ by $\Delta t$ and $V\times \Delta t$, respectively. 
After time delays, the nullifiers for the hexagonal clusters B with the label $k$ are described as
\begin{eqnarray}
\{&\hat{p}&_{{\rm B},1,k}+\hat{q}_{{\rm B},2,k+1}-\hat{q}_{{\rm B},6,k+V},\nonumber\\
&\hat{p}&_{{\rm B},2,k+1}+\hat{q}_{{\rm B},1,k}-\hat{q}_{{\rm B},3,k+N+1},\nonumber \\
&\hat{p}&_{{\rm B},3,k+N+1}-\hat{q}_{{\rm B},2,k+1}+\hat{q}_{{\rm B},4,k+N+1+V},\nonumber\\
&\hat{p}&_{{\rm B},4,k+N+1+V}+\hat{q}_{{\rm B},3,k+N+1}-\hat{q}_{{\rm B},5,k+N+V},\nonumber \\
&\hat{p}&_{{\rm B},5,k+N+V}-\hat{q}_{{\rm B},4,k+N+1+V}+\hat{q}_{{\rm B},6,k+V},\nonumber\\
&\hat{p}&_{{\rm B},6,k+V}-\hat{q}_{{\rm B},1,k}+\hat{q}_{{\rm B},5,k+N+V}\},
\end{eqnarray}  
whereas qumodes in the hexagonal cluster A maintain a time series, as shown in Fig. \ref{fig3}(v).
Then, a beam-splitter coupling between qumodes in the hexagonal clusters A and B with a same timing $T$ is implemented, as shown in Fig. \ref{fig2}(a)(vi) and (d). 
For lack of space, we only cover nullifiers for qumodes A$_{1,k}$ and A$_{2,k}$ in the following (see Appendix B for details on the transformation of nullifiers).
Nullifiers for qumodes A1 and A2 after a beam-splitter coupling with B1 and B2 are described as 
 \begin{eqnarray}
\{ &\hat{p}&_{{\rm A},1,k}+\hat{p}_{{\rm B},1,k}+\hat{q}_{{\rm A},2,k}+\hat{q}_{{\rm B},2,k}-\hat{q}_{{\rm A},6,k}-\hat{q}_{{\rm B},6,k},\nonumber\\
 &\hat{p}&_{{\rm A},2,k}+\hat{p}_{{\rm B},2,k}+\hat{q}_{{\rm A},1,k}+\hat{q}_{{\rm B},1,k}-\hat{q}_{{\rm A},3,k}-\hat{q}_{{\rm B},3,k}\},
\end{eqnarray}  
respectively.
By taking linear combinations and replacing labels, we obtain the nullifiers for qumodes A$_{1,k}$ and A$_{2,k}$ as below equations;
\begin{widetext}
\begin{eqnarray}
 \hat{p}_{{\rm A},1,k}&+&\frac{1}{2}(\hat{q}_{{\rm A},2,k}+\hat{q}_{{\rm B},2,k}+\hat{q}_{{\rm A},2,k+1}-\hat{q}_{{\rm B},2,k+1}-\hat{q}_{{\rm A},6,k}-\hat{q}_{{\rm B},6,k}-\hat{q}_{{\rm A},6,k+V}+\hat{q}_{{\rm B},6,k+V}), \nonumber \\
 \hat{p}_{{\rm A},2,k}&+&\frac{1}{2}(\hat{q}_{{\rm A},1,k}+\hat{q}_{{\rm B},1,k}+\hat{q}_{{\rm A},1,k-1}-\hat{q}_{{\rm B},1,k-1}-\hat{q}_{{\rm A},3,k}-\hat{q}_{{\rm B},3,k}-\hat{q}_{{\rm A},3,k+N}+\hat{q}_{{\rm B},3,k+N}).
\end{eqnarray}  
\end{widetext}
In a similar manner to the nullifiers for qumodes A$_{1,k}$ and A$_{2,k}$, we can obtain those for other qumodes.

\subsection{Verification of the generated topological cluster state}
\begin{figure}[b]
\begin{center}
 \includegraphics[angle=0, width=1.0\columnwidth]{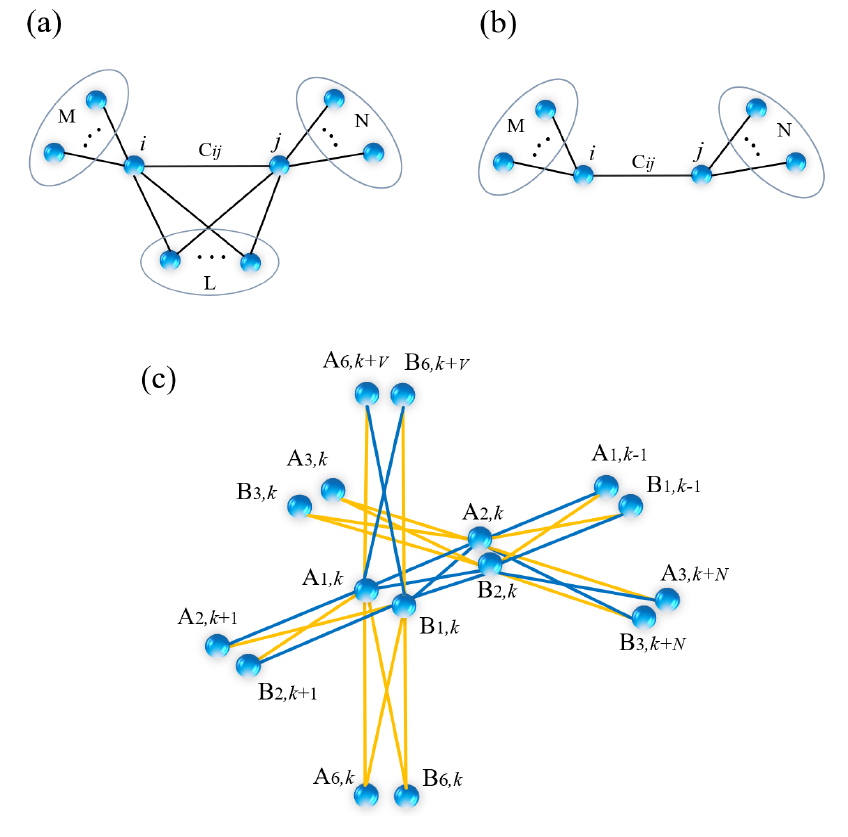} 
\caption{ Verification of the cluster sate. (a) Separability for a general cluster state. $C_{i,j}$ represents the edge-weight factor for qumodes $i$ and $j$. (b) Separability for  a particular cluster state. (c) Separability for the topological cluster state generated by using our method, focusing on qumodes ${\rm A}_{1,k}$ and ${\rm A}_{2,k}$.}
\label{fig4}
\end{center}
\end{figure}

We discuss sufficient conditions of entanglement for the generated cluster state by using the van Loock-Furusawa inseparability criteria \cite{Loock1} in order to verify the generated topological cluster state. 
Here we consider the $K$-mode cluster state for the general case.
The nullifiers for the general cluster state are given by $\delta =\hat{\bf p} -{\bf C} \hat{\bf x} $, where $\hat{\bf p}$ and $\hat{\bf x}$ are column vectors of momentum and position operators, respectively, and {\bf C} is an $K \times K$ weighted adjacency matrix \cite{Meni7}. The nullifiers for neighboring modes $i$ and $j$ are described as 
\begin{eqnarray}
\hat{\delta}_{i} &=& \hat{p}_{i}-C_{ij}\hat{q}_{j}-\sum_{m\in M}C_{im}\hat{q}_{m}-\sum_{l\in L}C_{il}\hat{q}_{l}, \nonumber \\
\hat{\delta}_{j} &=& \hat{p}_{j}-C_{ji}\hat{q}_{i}-\sum_{n\in N}C_{jn}\hat{q}_{n}-\sum_{l\in L}C_{jl}\hat{q}_{l},
\end{eqnarray}  
where $m$, $n$, and $l$ are labels for qumodes belonging to the multimode cluster states M, N, and L, as shown in Fig. \ref{fig4}(a).
We then consider the multimode cluster states M and N in Fig. \ref{fig4}(b) to deal with the multimode cluster state generated by our method, since the multimode cluster state L in Fig. \ref{fig4}(a) does not exist in our case.
In this case, nullifiers for neighboring modes $i$ and $j$ are described as 
\begin{eqnarray}
\hat{\delta}_{i} &=& \hat{p}_{i}-C_{ij}\hat{q}_{j}-\sum_{m\in M}C_{im}\hat{q}_{m}, \nonumber \\
\hat{\delta}_{j} &=& \hat{p}_{j}-C_{ji}\hat{q}_{i}-\sum_{n\in N}C_{jn}\hat{q}_{n},
\end{eqnarray}  
respectively. For the necessary condition of an inseparability between qumodes $i$ and $j$, if a quantum state is not separable into two subsets $S_{\alpha}$ and $S_{\beta}$, the inequality 
\begin{equation}
\langle \Delta^2 \hat{\delta}_{i}\rangle +\langle \Delta^2 \hat{\delta}_{j}\rangle < 2 {\hbar}|C_{ij} | \hspace{15pt} i\in S_{\alpha}, j \in S_{\beta}\label{inequal}
\end{equation}  
is satisfied, where the $S_{\alpha}$ and $ S_{\beta}$ are any bipartition of the set of all relevant qumodes. In Fig. \ref{fig4}(b), $S_{\alpha}$ is composed of the qumode $i$ and the multimode cluster state M, and $S_{\beta}$ is composed of the qumode $j$ and the multimode cluster states N.
For the necessary condition of an inseparability for the $K$-mode cluster state, if all inequalities for the nearest neighbor modes $i$ and $j$ in the $K$-mode cluster state are satisfied, the $K$-mode cluster state is fully entangled.

To obtain the necessary condition for our method, we see the qumodes ${\rm A}_{1,k}$ and ${\rm A}_{2,k}$ described in Fig. \ref{fig4}(c).
The nullifiers of  the qumodes ${\rm A}_{1,k}$ and ${\rm A}_{2,k}$ are described as 
\begin{eqnarray}
 \hat{\delta}_{{\rm A},1,k} =\hat{p}_{{\rm A},1,k}+\frac{1}{2}(\hat{q}_{{\rm A},2,k}+\hat{q}_{{\rm B},2,k}+\hat{q}_{{\rm A},2,k+1}-\hat{q}_{{\rm B},2,k+1}\nonumber\\
 -\hat{q}_{{\rm A},6,k}-\hat{q}_{{\rm B},6,k}-\hat{q}_{{\rm A},6,k+V}+\hat{q}_{{\rm B},6,k+V}), \label{null10}\\
  \hat{\delta}_{{\rm A},2,k} = \hat{p}_{{\rm A},2,k}+\frac{1}{2}(\hat{q}_{{\rm A},1,k}+\hat{q}_{{\rm B},1,k}+\hat{q}_{{\rm A},1,k-1}-\hat{q}_{{\rm A},1,k-1}\nonumber\\
 -\hat{q}_{{\rm A},3,k}+\hat{q}_{{\rm B},3,k}-\hat{q}_{{\rm A},3,k+N}+\hat{q}_{{\rm B},3,k+N}), \label{null2}
 \end{eqnarray} 
respectively.
We apply the generated cluster state with our method to Eqs. (\ref{inequal})-(\ref{null2}) as 
\begin{equation}
\langle \Delta^2 \hat{\delta}_{{\rm A},1,k}\rangle +\langle \Delta^2  \hat{\delta}_{{\rm A},2,k}\rangle =3\hbar {\rm e}^{-2r}< {\hbar},
\end{equation}  
where we use Eq. (12), e.g. the variance for qumodes,
\begin{eqnarray}
&\langle(&\hat{p}_{{\rm A(B)},2n-1,k})^2\rangle={{\rm e}^{-2r}}\langle(\hat{p}_{{\rm A(B)},2n-1,k}^{(0)})^2\rangle=\frac{{\rm e}^{-2r}}{2}, \nonumber \\
&\langle(&\hat{q}_{{\rm A(B)},2n,k})^2\rangle={{\rm e}^{-2r}}\langle(\hat{q}_{{\rm A(B)},2n,k}^{(0)})^2\rangle=\frac{{\rm e}^{-2r}}{2}.
\end{eqnarray}
 (see Appendix C for details on the calculation for Eq. (31)).
Thus, we can verify the generation of the topological cluster state, if the inequality 
\begin{equation}
{\rm e}^{-2r} < \frac{1}{3}
\end{equation}
is satisfied.
From the van-Loock-Furusawa criterion \cite{Loock1}, the required squeezing level to satisfy the above inequality is $\sim$-4.77dB.
Consequently, our method provides almost the same required squeezing level, -4.5 dB, to show sufficient conditions of entanglement for the 2-dimensional cluster state which has been demonstrated in Ref. \cite{Warit}.

Here we mention that this benefit of the feasible squeezing of the generated cluster state comes from the economical use of a beam-splitter coupling. 
Generally, a beam-splitter coupling leads to a decrease in the amplitude of the edge-weight factor \cite{Meni5}, without the aid of the decomposition technique in Ref. \cite{Loock}.
Besides, the smaller the amplitude of the edge-weight factor, the more the required squeezing level to show sufficient conditions is \cite{Yoko,Warit}.
In our method, we firstly generate appropriate small-scale building blocks, i.e., hexagonal cluster states by using the decomposition technique. Then the topological cluster state is constructed from building blocks by using the only one beam-splitter coupling per node of the topological cluster state.
In the conventional method, on the other hand, a topological cluster state will be generated from the building blocks, which is two-mode entangled states, by using the more than three beam-splitter couplings per node.
Hence, our method can provide the feasible squeezing to verify the generated cluster state.

\subsection{Robustness against analog errors}
In QC with squeezed vacuum states, the displacement errors derived from a finite squeezing generally propagate between qumodes by two-qubit gates, and are accumulated due to the quantum-teleportation-based gate in MBQC. 
Thus, the quantum error correction is needed to correct them for implementing large-scale quantum computation by using an appropriate code such as the GKP qubit \cite{GKP}.
Nevertheless, the large displacement error occurs as the qubit-level error, i.e., bit- and phase-flip errors in the code word of the GKP qubit.
Thus, the accumulation of displacement errors should be reduced to improve the noise tolerance against analog errors. 
In this subsection, we show the second advantage of our approach, i.e.,  a desirable noise tolerance against analog errors during MBQC.

\begin{figure}[t]
\begin{center}
 \includegraphics[angle=0, width=1.0\columnwidth]{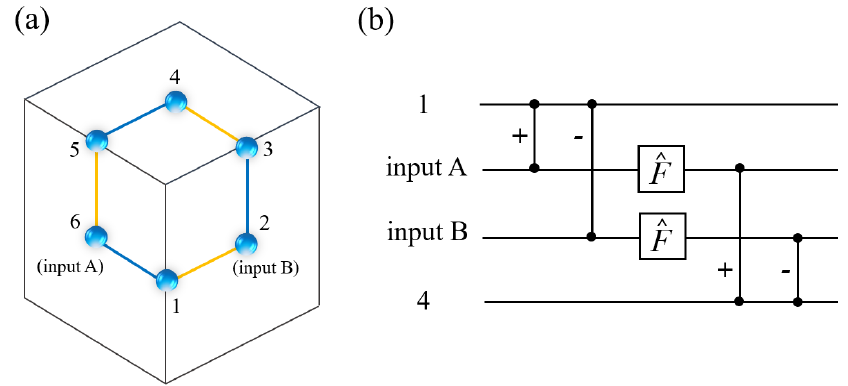} 
\caption{ Error propagation in the generated topological cluster state. (a) Error propagation from the qumode 1 to the qumode 4 , where qumodes 2 and 6 are input states, and qumodes 1 and 4 are used for the syndrome measurement of $Z$ and $X$ stabilizers. (b) An equivalent quantum circuit for MBQC on the cluster state within the framework for a circuit-based model. $\hat{F}$ denotes the Fourier transformation and is implemented by the measurement of the qumode in the $p$ quadrature. $\pm$ denotes the sign of interaction strength of the CZ gate, i.e., the sign of the edge-weight factor.}
\label{fig5}
\end{center}
\end{figure}
In the following, let us look the noise propagation between squeezed vacuum states, since the detailed analysis of the quantum error correction with the GKP qubit is out of the scope of the present work.
For simplicity, we focus on the propagation of the displacement error from the qumode 1 to the qumode 4, as shown in Fig. \ref{fig5}(a), assuming that the qumodes 1 and 4 are measured in the $p$ quadrature for the $Z$ and $X$ stabilizers, respectively.
Fig. \ref{fig5}(b) shows an equivalent circuit for MBQC on the cluster state.
We here introduce the CZ gate which corresponds to the operator exp(-$i g\hat{q}_{\rm j}\hat{q}_{\rm k}$) for qumodes $j$ and $k$ with the factor $g$ corresponding to the magnitude of the edge-weight factor.
The CZ gate transforms displacement errors in the $p$ quadrature as
\begin{equation}
\overline{\Delta}_{\rm {\it p},j} \to \overline{\Delta}_{\rm {\it p},j}-g \overline{\Delta}_{\rm {\it q},k}, \hspace{10pt}
\overline{\Delta}_{\rm {\it p},k} \to \overline{\Delta}_{\rm {\it p},k}-g \overline{\Delta}_{\rm {\it q},j},
\end{equation}
where $\overline{\Delta}_{\rm {\it q},j} ( \overline{\Delta}_{\rm {\it p},j})$ and $\overline{\Delta}_{\rm {\it q},k}  (\overline{\Delta}_{\rm {\it p},k} )$ are values of the displacement error for qumodes $j$ and $k$ in the $q(p)$ quadrature, respectively. 
Let us consider only the displacement error of qumode 1 in the $q$ quadrature, $\overline{\Delta}_{q,1}$; the deviation errors of qumodes except for the qumode 1 are zero.
Taking into account the CZ gate, the deviation errors of qumodes 2 and 6 in the $p$ quadrature are described as
\begin{equation}
\overline{\Delta}_{p,2}=g\overline{\Delta}_{q,1}, \hspace{10pt}
\overline{\Delta}_{p,6}=-g\overline{\Delta}_{q,1}.
\end{equation}
After the measurement on qumodes 2 and 6 in the $p$ quadrature, displacement errors of the qumodes 2 and 6 in the $p$ quadrature are transformed to those of the qumodes of 3 and 5 in the $q$ quadrature as 
\begin{equation}
\overline{\Delta}_{q,3}= \overline{\Delta}_{q,1}, \hspace{10pt}
\overline{\Delta}_{q,5}=\overline{\Delta}_{q,1}.
\end{equation}
We note that the displacement errors are amplified by $g$, according to the procedure of MBQC.
Those of deviation errors of qumodes 3 and 5 eventually propagate on the qumode 4 in the $p$ quadrature by the CZ gates.
This transformation corresponds to the Fourier transformation on the inputs A and B in Fig. \ref{fig5}(b) within the framework for a circuit-based model.
After the CZ gates between qumodes 3 and 4, and 5 and 4, the deviation errors of the qumode 4 is 
\begin{equation}
\overline{\Delta}_{q,4}=g\overline{\Delta}_{q,1}-g\overline{\Delta}_{q,1}=0, 
\end{equation}
where the edge-weight factors with respect to the qumodes 3 and 5 are $+g$ and $-g$, respectively.
We can see that the analog error derived from the qumode 1 is canceled out in the qumode 4, and therefore the generated topological cluster state has a robustness against displacement errors during topologically protected MBQC \cite{Note2}.
Since this feature is obtained thanks to the sign of the edge-weight factors of the generated topological cluster state, our method is practical to realize fault-tolerant MBQC with the robustness of analog errors, in addition to a reasonable squeezing level for the verification of the entanglement.

In addition, we note the effect of the edge-weight factor on the quantum error correction with the GKP qubit.
To perform the quantum error correction with the GKP qubit, the amplitude of edge-weight factors should be set to 1, since the amplitude of the interaction of the two-qubit gate between GKP qubits should be 1 in the code word of the GKP qubit.
Therefore, the strength of the entanglement of the topological cluster state will be recovered to adjust the amplitude of the edge-weight factor to 1 \cite{Glan, Wan}.
As a result, this entanglement recovery increases the noise derived from a finite squeezing of the squeezed vacuum states by the inverse of the edge-weight factor.
For example, the amplitude of the edge-weight factor of the 3-dimensional cluster state by using only the time-domain multiplexing approach is 1/${4\sqrt{2}}$ \cite{Wu}.
Thus, our method with the amplitude of the edge-weight factor 1/2 has an advantage for performing quantum error correction with the GKP qubit \cite{Note3}.

\section{Discussion and conclusion}\label{Sec5}
In this work, we have proposed the method to generate the topological cluster state for implementing topologically-protected MBQC with the linear optics.
Our method makes effective use of the advantage of the time-domain multiplexing approach which is currently a promising way to realize large-scale MBQC among various approaches and physical systems, such as a superconducting and an ion-trap, due to the ability to generate the large-scale cluster state.
In our method, the squeezing level required for verifying the generated cluster state is an experimentally feasible value, $\sim$-4.77dB, which is almost the same level with the 2-dimensional cluster state generated by using the conventional method, -4.5dB \cite{Warit}.
Moreover, in the generated cluster state, analog errors are canceled out and prevented from propagating between the qumodes thanks to the feature of a sign of an edge-weight factor. 
For the quantum error correction with the GKP qubit, the generated cluster state has an advantage due to the smaller amplitude of the edge-weight factor, compared to that by using only the time-domain multiplexing approach.
These features are compatible with the analog quantum error correction \cite{KF1} and high-threshold topologically protected MBQC with the GKP qubit \cite{KF2,KF4}. 
High-threshold topologically protected MBQC on the topological cluster state generated by our method will provide a new approach to implement large-scale MBQC with an experimentally feasible squeezing level.
In addition, we mention the resource usage for the cubic phase gate to implement one-mode non-Gaussian operation for universality. In our setup, the cubic phase gate can be implemented by injecting the cubic phase state into the cluster state, as discussed in Ref. \cite{Warit}.
In future work we will investigate the resource usage such as the GKP qubit and the cubic phase state with our method.
Lastly, although we apply our approach to the topological cluster state in this paper, our method can be applied to a variety of entangled states such as the 3-dimensional lattice for a color code \cite{Bomb, Brown}, the 2-dimensional honeycomb state \cite{Nest}, and so on.
Furthermore, our method can be applied to several promising architectures for a scalable quantum circuit \cite{Takeda, Raf, Takeda2}.
Hence, we believe this work will provide a new way to generate the large-scale resource state to implement fault-tolerant MBQC with continuous variables.

\section*{Acknowledgements}
This work was partly supported by Japan Society for the Promotion of Science (JSPS) KAKENHI (grant 18H05207), CREST (Grant No. JP- MJCR15N5), UTokyo Foundation, and donations from Nichia Corporation.
W. A. acknowledges financial support from the Japan Society for the Promotion of Science (JSPS).

\begin{widetext}
\section*{Appendix A: Calculation of nullifiers for the 1-dimensional cluster state}
In the following we describe how to calculate the nullifiers and be used for the verification. The initial nullifiers for qumodes A and B in the temporal mode index $k$ are defined as
\begin{equation}
\{ \hat{q}_{{\rm A},k},\hspace{3pt}\hat{p}_{{\rm A},k}\},
\tag{A1}
\end{equation}
since qumodes A and B from OPOs have the position and momentum squeezing, respectively.
The nullifiers after the first beam-splitter coupling become 
\begin{equation}
\{\hat{q}_{{\rm A},k}+\hat{q}_{{\rm B},k},-\hat{p}_{{\rm A},k}+\hat{p}_{{\rm B},k} \}.
\tag{A2}
\end{equation}  
Then the time delay on qumodes B transforms the nullifiers as
\begin{equation}
\{\hat{q}_{{\rm A},k}+\hat{q}_{{\rm B},k+1},-\hat{p}_{{\rm A},k}+\hat{p}_{{\rm B},k+1} \}.
\tag{A3}
\end{equation} 
After the second beam-splitter coupling, we obtain nullifiers
\begin{equation}
\{\hat{q}_{{\rm A},k}+\hat{q}_{{\rm B},k}-\hat{q}_{{\rm A},k+1}+\hat{q}_{{\rm B},k+1},
-\hat{p}_{{\rm A},k}-\hat{p}_{{\rm B},k}-\hat{p}_{{\rm A},k+1}+\hat{p}_{{\rm B},k+1} \}.
\tag{A4}
\end{equation} 
From $\hat{q}_{{\rm A(B)},k}=\hat{q}^{({\rm ii})}_{{\rm A(B)},k}$ and $\hat{p}_{{\rm A(B)},k}=\hat{p}^{({\rm ii})}_{{\rm A(B)},k},$ the nullifiers of mode $k$ for the generated 1-dimensional cluster state in the $q$ and $p$ operators, $\hat{\delta}_{k}^{q}$ and  $\hat{\delta}_{k}^{p}$, are obtained as
\begin{align}
\hat{\delta}_{k}^{q}=\hat{q}^{({\rm ii})}_{{\rm A},k}+\hat{q}^{({\rm ii})}_{{\rm B},k}-\hat{q}^{({\rm ii})}_{{\rm A},k+1}+\hat{q}^{({\rm ii})}_{{\rm B},k+1}, \hspace{10pt}
\hat{\delta}_{k}^{p}=-\hat{p}^{({\rm ii})}_{{\rm A},k}-\hat{p}^{({\rm ii})}_{{\rm B},k}-\hat{p}^{({\rm ii})}_{{\rm A},k+1}+\hat{p}^{({\rm ii})}_{{\rm B},k+1},
\tag{A5}
\end{align}
respectively, as described in Eqs. (5) and (6) in the main text.

We here give another description of nullifiers for the generated 1-dimensional cluster state in order to characterize the color of the link corresponding to the sign of edge-weight factors for the generated state. 
Since linear combinations of the nullifiers are also nullifiers because of the property of the nullifier, we obtain the nullifiers for the generated 1-dimensional cluster state by taking linear combinations of them as
\begin{align}
\{&\hat{q}_{{\rm A},k}-\frac{1}{2}(\hat{q}_{{\rm A},k+1}-\hat{q}_{{\rm B},k+1}+\hat{q}_{{\rm A},k-1}+\hat{q}_{{\rm B},k-1}), \hspace{10pt}
\hat{q}_{{\rm B},k}-\frac{1}{2}(\hat{q}_{{\rm A},k+1}-\hat{q}_{{\rm B},k+1}-\hat{q}_{{\rm A},k-1}-\hat{q}_{{\rm B},k-1}), \nonumber \\
&\hat{p}_{{\rm A},k}+\frac{1}{2}(\hat{p}_{{\rm A},k+1}-\hat{p}_{{\rm A},k-1}+\hat{p}_{{\rm B},k+1}+\hat{p}_{{\rm B},k-1}), \hspace{10pt}
\hat{p}_{{\rm B},k}+\frac{1}{2}(\hat{p}_{{\rm A},k+1}-\hat{p}_{{\rm A},k-1}-\hat{p}_{{\rm B},k+1}-\hat{p}_{{\rm B},k-1})
\}.
\tag{A6}
\end{align} 
Considering that nullifiers for the generated cluster state are given by $\{\hat{q}_{{\rm A(B)}, k} -{C}_{kj}\hspace{3pt}  \hat{q}_{{\rm A(B)}, j},\hspace{5pt} \hat{p}_{{\rm A(B)}, k} +{C}_{kj}\hspace{3pt} \hat{p}_{{\rm A(B)}, j}\}$, we obtain the weights of the generated cluster state, ${C}_{kj}$, where $j$ is the neighboring qumodes of the qumode $k$.
The color of the generated cluster is determined by the sign of weights, i.e., the blue and yellow edges represent + and - signs, respectively, as shown in Fig.1 in the main text.

\section*{Appendix B: Nullifiers for the generated topological cluster state}
We describe the transformation of nullifiers through the beam-splitter coupling between qumodes in the hexagonal clusters A and B, as shown in Fig. \ref{fig2}(vi) in the main text. 
After the beam-splitter coupling, nullifiers for the hexagonal cluster state A become
 \begin{align}
\{ &\hat{p}_{{\rm A},1,k}+\hat{p}_{{\rm B},1,k}+\hat{q}_{{\rm A},2,k}+\hat{q}_{{\rm B},2,k}-\hat{q}_{{\rm A},6,k}-\hat{q}_{{\rm B},6,k},\nonumber\\
 &\hat{p}_{{\rm A},2,k}+\hat{p}_{{\rm B},2,k}+\hat{q}_{{\rm A},1,k}+\hat{q}_{{\rm B},1,k}-\hat{q}_{{\rm A},3,k}-\hat{q}_{{\rm B},3,k},\nonumber\\
 &\hat{p}_{{\rm A},3,k}+\hat{p}_{{\rm B},3,k}-\hat{q}_{{\rm A},2,k}-\hat{q}_{{\rm B},2,k}+\hat{q}_{{\rm A},4,k}+\hat{q}_{{\rm B},4,k},\nonumber\\
 &\hat{p}_{{\rm A},4,k}+\hat{p}_{{\rm B},4,k}+\hat{q}_{{\rm A},3,k}+\hat{q}_{{\rm B},3,k}-\hat{q}_{{\rm A},5,k}-\hat{q}_{{\rm B},5,k},\nonumber\\
 &\hat{p}_{{\rm A},5,k}+\hat{p}_{{\rm B},5,k}-\hat{q}_{{\rm A},4,k}-\hat{q}_{{\rm B},4,k}+\hat{q}_{{\rm A},6,k}+\hat{q}_{{\rm B},6,k},\nonumber\\
 &\hat{p}_{{\rm A},6,k}+\hat{p}_{{\rm B},6,k}-\hat{q}_{{\rm A},1,k}-\hat{q}_{{\rm B},1,k}+\hat{q}_{{\rm A},5,k}+\hat{q}_{{\rm B},5,k},\}.
\tag{B1}
\end{align}  
Nullifiers for the hexagonal cluster state B become
 \begin{align}
\{ &\hat{p}_{{\rm A},1,k}-\hat{p}_{{\rm B},1,k}+\hat{q}_{{\rm A},2,k+1}-\hat{q}_{{\rm B},2,k+1}-\hat{q}_{{\rm A},6,k+V}+\hat{q}_{{\rm B},6,k+V},\nonumber\\
&\hat{p}_{{\rm A},2,k}-\hat{p}_{{\rm B},2,k}+\hat{q}_{{\rm A},1,k-1}-\hat{q}_{{\rm B},1,k-1}-\hat{q}_{{\rm A},3,k+N}+\hat{q}_{{\rm B},3,k+N},\nonumber\\
 &\hat{p}_{{\rm A},3,k}-\hat{p}_{{\rm B},3,k}-\hat{q}_{{\rm A},2,k-N}+\hat{q}_{{\rm B},2,k-N}+\hat{q}_{{\rm A},4,k+V}-\hat{q}_{{\rm B},4,k+V},\nonumber\\
 &\hat{p}_{{\rm A},4,k}-\hat{p}_{{\rm B},4,k}+\hat{q}_{{\rm A},3,k-V}-\hat{q}_{{\rm B},3,k-V}-\hat{q}_{{\rm A},5,k-1}+\hat{q}_{{\rm B},5,k-1},\nonumber\\
 &\hat{p}_{{\rm A},5,k}-\hat{p}_{{\rm B},5,k}-\hat{q}_{{\rm A},4,k+1}+\hat{q}_{{\rm B},4,k+1}+\hat{q}_{{\rm A},6,k-N}-\hat{q}_{{\rm B},6,k-N},\nonumber\\
 &\hat{p}_{{\rm A},6,k}-\hat{p}_{{\rm B},6,k}-\hat{q}_{{\rm A},1,k-V}+\hat{q}_{{\rm B},1,k-V}+\hat{q}_{{\rm A},5,k+N}-\hat{q}_{{\rm B},5,k+N}
\}.
\tag{B2}
\end{align}  
By taking linear combinations and replacing labels, we obtain the nullifiers for qumodes A as
\begin{align}
& \hat{\delta}_{{\rm A},1,k} = \hat{p}_{{\rm A},1,k}+\frac{1}{2}(\hat{q}_{{\rm A},2,k}+\hat{q}_{{\rm B},2,k}+\hat{q}_{{\rm A},2,k+1}-\hat{q}_{{\rm B},2,k+1}-\hat{q}_{{\rm A},6,k}-\hat{q}_{{\rm B},6,k}-\hat{q}_{{\rm A},6,k+V}+\hat{q}_{{\rm B},6,k+V}), \nonumber \\
&\hat{\delta}_{{\rm A},2,k} = \hat{p}_{{\rm A},2,k}+\frac{1}{2}(\hat{q}_{{\rm A},1,k}+\hat{q}_{{\rm B},1,k}+\hat{q}_{{\rm A},1,k-1}-\hat{q}_{{\rm B},1,k-1}-\hat{q}_{{\rm A},3,k}-\hat{q}_{{\rm B},3,k}-\hat{q}_{{\rm A},3,k+N}+\hat{q}_{{\rm B},3,k+N}), \nonumber \\
&\hat{\delta}_{{\rm A},3,k} = \hat{p}_{{\rm A},3,k}+\frac{1}{2}(-\hat{q}_{{\rm A},2,k}-\hat{q}_{{\rm B},2,k}-\hat{q}_{{\rm A},2,k-N}+\hat{q}_{{\rm B},2,k-N}+\hat{q}_{{\rm A},4,k}+\hat{q}_{{\rm B},4,k}+\hat{q}_{{\rm A},4,k+V}-\hat{q}_{{\rm B},4,k+V}), \nonumber \\
&\hat{\delta}_{{\rm A},4,k} = \hat{p}_{{\rm A},4,k}+\frac{1}{2}(\hat{q}_{{\rm A},3,k}+\hat{q}_{{\rm B},3,k}+\hat{q}_{{\rm A},3,k-V}-\hat{q}_{{\rm B},3,k-V}-\hat{q}_{{\rm A},5,k}-\hat{q}_{{\rm B},5,k}-\hat{q}_{{\rm A},5,k-1}+\hat{q}_{{\rm B},5,k-1}), \nonumber \\
&\hat{\delta}_{{\rm A},5,k} = \hat{p}_{{\rm A},5,k}+\frac{1}{2}(-\hat{q}_{{\rm A},4,k}-\hat{q}_{{\rm B},4,k}-\hat{q}_{{\rm A},4,k+1}+\hat{q}_{{\rm B},4,k+1}+\hat{q}_{{\rm A},6,k}+\hat{q}_{{\rm B},6,k}+\hat{q}_{{\rm A},6,k-N}-\hat{q}_{{\rm B},6,k-N}), \nonumber \\
&\hat{\delta}_{{\rm A},6,k} = \hat{p}_{{\rm A},6,k}+\frac{1}{2}(-\hat{q}_{{\rm A},1,k}-\hat{q}_{{\rm B},1,k}-\hat{q}_{{\rm A},1,k-V}+\hat{q}_{{\rm B},1,k-V}+\hat{q}_{{\rm A},5,k}+\hat{q}_{{\rm B},5,k}+\hat{q}_{{\rm A},5,k+N}-\hat{q}_{{\rm B},5,k+N}), \tag{B3}
\end{align}  
and obtain the nullifiers for qumodes B as
\begin{align}
&\hat{\delta}_{{\rm B},1,k} = \hat{p}_{{\rm B},1,k}+\frac{1}{2}(\hat{q}_{{\rm A},2,k}+\hat{q}_{{\rm B},2,k}-\hat{q}_{{\rm A},2,k+1}+\hat{q}_{{\rm B},2,k+1}-\hat{q}_{{\rm A},6,k}-\hat{q}_{{\rm B},6,k}+\hat{q}_{{\rm A},6,k+V}-\hat{q}_{{\rm B},6,k+V}), \nonumber \\
&\hat{\delta}_{{\rm B},2,k} =\hat{p}_{{\rm B},2,k}+\frac{1}{2}(\hat{q}_{{\rm A},1,k}+\hat{q}_{{\rm B},1,k}-\hat{q}_{{\rm A},1,k-1}+\hat{q}_{{\rm B},1,k-1}-\hat{q}_{{\rm A},3,k}-\hat{q}_{{\rm B},3,k}+\hat{q}_{{\rm A},3,k+N}-\hat{q}_{{\rm B},3,k+N}), \nonumber \\
&\hat{\delta}_{{\rm B},3,k} =\hat{p}_{{\rm B},3,k}+\frac{1}{2}(-\hat{q}_{{\rm A},2,k}-\hat{q}_{{\rm B},2,k}+\hat{q}_{{\rm A},2,k-N}-\hat{q}_{{\rm B},2,k-N}+\hat{q}_{{\rm A},4,k}+\hat{q}_{{\rm B},4,k}-\hat{q}_{{\rm A},4,k+V}+\hat{q}_{{\rm B},4,k+V}), \nonumber \\
&\hat{\delta}_{{\rm B},4,k} =\hat{p}_{{\rm B},4,k}+\frac{1}{2}(\hat{q}_{{\rm A},3,k}+\hat{q}_{{\rm B},3,k}-\hat{q}_{{\rm A},3,k-V}+\hat{q}_{{\rm B},3,k-V}-\hat{q}_{{\rm A},5,k}-\hat{q}_{{\rm B},5,k}+\hat{q}_{{\rm A},5,k-1}-\hat{q}_{{\rm B},5,k-1}), \nonumber \\
&\hat{\delta}_{{\rm B},5,k} =\hat{p}_{{\rm B},5,k}+\frac{1}{2}(-\hat{q}_{{\rm A},4,k}-\hat{q}_{{\rm B},4,k}+\hat{q}_{{\rm A},4,k+1}-\hat{q}_{{\rm B},4,k+1}+\hat{q}_{{\rm A},6,k}+\hat{q}_{{\rm B},6,k}-\hat{q}_{{\rm A},6,k-N}+\hat{q}_{{\rm B},6,k-N}), \nonumber \\
&\hat{\delta}_{{\rm B},6,k} =\hat{p}_{{\rm B},6,k}+\frac{1}{2}(-\hat{q}_{{\rm A},1,k}-\hat{q}_{{\rm B},1,k}+\hat{q}_{{\rm A},1,k-V}-\hat{q}_{{\rm B},1,k-V}+\hat{q}_{{\rm A},5,k}+\hat{q}_{{\rm B},5,k}-\hat{q}_{{\rm A},5,k+N}+\hat{q}_{{\rm B},5,k+N}). \tag{B4}
\end{align}

\section*{Appendix C: Calculation of the inequality for the generated topological cluster state}
We explain the calculation in Eq. (31) in the main text. 
Using Eqs. (12)-(15) in the main text, the operators for qumodes in the hexagonal cluster state A are represented by
\begin{align}
&\hat{a}^{({\rm iii}) }_{{\rm A}{,1,k}}= i\hat{a}^{({\rm ii}) }_{{\rm A}{,1,k}}=\frac{i}{\sqrt{2}}(\hat{a}^{({\rm i}) }_{{\rm A}{,1,k}}- \hat{a}^{({\rm i}) }_{{\rm A}{,4,k}})=\frac{i}{2}(\hat{a}^{({\rm 0}) }_{{\rm A}{,1,k}}- \hat{a}^{({\rm 0}) }_{{\rm A}{,6,k}}-\hat{a}^{({\rm 0}) }_{{\rm A}{,5,k}}-\hat{a}^{({\rm 0}) }_{{\rm A}{,4,k}}), \nonumber \\
&\hat{a}^{({\rm iii}) }_{{\rm A}{,2,k}}=\hat{a}^{({\rm ii}) }_{{\rm A}{,2,k}}=\frac{1}{\sqrt{2}}(\hat{a}^{({\rm i}) }_{{\rm A}{,5,k}}+ \hat{a}^{({\rm i}) }_{{\rm A}{,2,k}})=\frac{1}{2}(\hat{a}^{({\rm 0}) }_{{\rm A}{,5,k}}- \hat{a}^{({\rm 0}) }_{{\rm A}{,4,k}}+\hat{a}^{({\rm 0}) }_{{\rm A}{,3,k}}+\hat{a}^{({\rm 0}) }_{{\rm A}{,2,k}}), \nonumber \\
&\hat{a}^{({\rm iii}) }_{{\rm A}{,3,k}}=i\hat{a}^{({\rm ii}) }_{{\rm A}{,3,k}}=\frac{i}{\sqrt{2}}(\hat{a}^{({\rm i}) }_{{\rm A}{,3,k}}- \hat{a}^{({\rm i}) }_{{\rm A}{,6,k}})=\frac{i}{2}(\hat{a}^{({\rm 0}) }_{{\rm A}{,3,k}}- \hat{a}^{({\rm 0}) }_{{\rm A}{,2,k}}-\hat{a}^{({\rm 0}) }_{{\rm A}{,1,k}}-\hat{a}^{({\rm 0}) }_{{\rm A}{,6,k}}), \nonumber \\
&\hat{a}^{({\rm iii}) }_{{\rm A}{,4,k}}= \hat{a}^{({\rm ii}) }_{{\rm A}{,4,k}}=\frac{1}{\sqrt{2}}(\hat{a}^{({\rm i}) }_{{\rm A}{,1,k}}+ \hat{a}^{({\rm i}) }_{{\rm A}{,4,k}})=\frac{1}{2}(\hat{a}^{({\rm 0}) }_{{\rm A}{,1,k}}- \hat{a}^{({\rm 0}) }_{{\rm A}{,6,k}}+\hat{a}^{({\rm 0}) }_{{\rm A}{,5,k}}+\hat{a}^{({\rm 0}) }_{{\rm A}{,4,k}}), \nonumber \\
&\hat{a}^{({\rm iii}) }_{{\rm A}{,5,k}}=i\hat{a}^{({\rm ii}) }_{{\rm A}{,5,k}}=\frac{i}{\sqrt{2}}(\hat{a}^{({\rm i}) }_{{\rm A}{,5,k}}- \hat{a}^{({\rm i}) }_{{\rm A}{,2,k}})=\frac{i}{2}(\hat{a}^{({\rm 0}) }_{{\rm A}{,5,k}}- \hat{a}^{({\rm 0}) }_{{\rm A}{,4,k}}-\hat{a}^{({\rm 0}) }_{{\rm A}{,3,k}}-\hat{a}^{({\rm 0}) }_{{\rm A}{,2,k}}), \nonumber \\
&\hat{a}^{({\rm iii}) }_{{\rm A}{,6,k}}=\hat{a}^{({\rm ii}) }_{{\rm A}{,6,k}}=\frac{1}{\sqrt{2}}(\hat{a}^{({\rm i}) }_{{\rm A}{,3,k}}+ \hat{a}^{({\rm i}) }_{{\rm A}{,6,k}})=\frac{1}{2}(\hat{a}^{({\rm 0}) }_{{\rm A}{,3,k}}- \hat{a}^{({\rm 0}) }_{{\rm A}{,2,k}}+\hat{a}^{({\rm 0}) }_{{\rm A}{,1,k}}+\hat{a}^{({\rm 0}) }_{{\rm A}{,6,k}}),
\tag{C1}
\end{align}
respectively. 
The operators for qumodes in the hexagonal cluster state B are derived in the same form as Eq. (C1).
After time delays on qumodes B, a beam-splitter coupling between qumodes A and B with a same timing $T$ is performed, as described in Eq. (17) in the main text. 
After the beam-splitter coupling, we obtain annihilation operators for A and B, $\hat{a}_{{\rm A}{,n,k}}$ and $\hat{a}_{{\rm B}{,n,k}}$ with $n=1,2, \cdots 6$, as
\begin{align}
&\hat{a}_{{\rm A}{,1,k}}=\hat{a}^{({\rm iii}) }_{{\rm A}{,1,k}}+\hat{a}^{({\rm iii}) }_{{\rm B}{,1,k}}, \hspace{15pt}
\hat{a}_{{\rm A}{,2,k}}=\hat{a}^{({\rm iii}) }_{{\rm A}{,2,k}}+\hat{a}^{({\rm iii}) }_{{\rm B}{,2,k-1}}, \hspace{15pt}
\hat{a}_{{\rm A}{,3,k}}=\hat{a}^{({\rm iii}) }_{{\rm A}{,3,k}}+\hat{a}^{({\rm iii}) }_{{\rm B}{,3,k-N-1}},\nonumber \\
&\hat{a}_{{\rm A}{,4,k}}=\hat{a}^{({\rm iii}) }_{{\rm A}{,4,k}}+\hat{a}^{({\rm iii}) }_{{\rm B}{,4,k-N-V-1}}, \hspace{15pt}
\hat{a}_{{\rm A}{,5,k}}=\hat{a}^{({\rm iii}) }_{{\rm A}{,5,k}}+\hat{a}^{({\rm iii}) }_{{\rm B}{,5,k-N-V}}, \hspace{15pt}
\hat{a}_{{\rm A}{,6,k}}=\hat{a}^{({\rm iii}) }_{{\rm A}{,6,k}}+\hat{a}^{({\rm iii}) }_{{\rm B}{,6,k-V}},
\tag{C2}
\end{align}
and
\begin{align}
&\hat{a}_{{\rm B}{,1,k}}=\hat{a}^{({\rm iii}) }_{{\rm A}{,1,k}}-\hat{a}^{({\rm iii}) }_{{\rm B}{,1,k}}, \hspace{15pt}
\hat{a}_{{\rm B}{,2,k}}=\hat{a}^{({\rm iii}) }_{{\rm A}{,2,k}}-\hat{a}^{({\rm iii}) }_{{\rm B}{,2,k-1}}, \hspace{15pt}
\hat{a}_{{\rm B}{,3,k}}=\hat{a}^{({\rm iii}) }_{{\rm A}{,3,k}}-\hat{a}^{({\rm iii}) }_{{\rm B}{,3,k-N-1}},\nonumber \\
&\hat{a}_{{\rm B}{,4,k}}=\hat{a}^{({\rm iii}) }_{{\rm A}{,4,k}}-\hat{a}^{({\rm iii}) }_{{\rm B}{,4,k-N-V-1}}, \hspace{15pt}
\hat{a}_{{\rm B}{,5,k}}=\hat{a}^{({\rm iii}) }_{{\rm A}{,5,k}}-\hat{a}^{({\rm iii}) }_{{\rm B}{,5,k-N-V}}, \hspace{15pt}
\hat{a}_{{\rm B}{,6,k}}=\hat{a}^{({\rm iii}) }_{{\rm A}{,6,k}}-\hat{a}^{({\rm iii}) }_{{\rm B}{,6,k-V}},
\tag{C3}
\end{align}
respectively. 
Using Eqs. (C2), (C3), and (12) in the main text, the nullifiers for qumodes A and B are obtained as
\begin{align}
&\hat{\delta}_{{\rm A},1,k} =\frac{{\rm e}^{-r}}{\sqrt{2}}(-\hat{q}^{(0)}_{{\rm A},2, k}-\hat{q}^{(0)}_{{\rm B},2, k}-\hat{q}^{(0)}_{{\rm A},4, k}-\hat{q}^{(0)}_{{\rm B},4, k}-\hat{q}^{(0)}_{{\rm A},6, k}-\hat{q}^{(0)}_{{\rm B},6, k}), \nonumber \\
&\hat{\delta}_{{\rm A},2,k} =\frac{{\rm e}^{-r}}{\sqrt{2}}(\hat{p}^{(0)}_{{\rm A},1, k}+\hat{p}^{(0)}_{{\rm B},1, k-1}+\hat{p}^{(0)}_{{\rm A},3, k}+\hat{p}^{(0)}_{{\rm B},3, k-1}+\hat{p}^{(0)}_{{\rm A},5, k}+\hat{p}^{(0)}_{{\rm B},5, k-1}), \nonumber \\
&\hat{\delta}_{{\rm A},3,k} =\frac{{\rm e}^{-r}}{\sqrt{2}}(-\hat{q}^{(0)}_{{\rm A},2, k}-\hat{q}^{(0)}_{{\rm B},2, k-N-1}+\hat{q}^{(0)}_{{\rm A},4, k}+\hat{q}^{(0)}_{{\rm B},4, k-N-1}-\hat{q}^{(0)}_{{\rm A},6, k}-\hat{q}^{(0)}_{{\rm B},6, k-N-1}), \nonumber \\
&\hat{\delta}_{{\rm A},4,k} =\frac{{\rm e}^{-r}}{\sqrt{2}}(\hat{p}^{(0)}_{{\rm A},1, k}+\hat{p}^{(0)}_{{\rm B},1, k-N-V-1}+\hat{p}^{(0)}_{{\rm A},3, k}+\hat{p}^{(0)}_{{\rm B},3, k-N-V-1}+\hat{p}^{(0)}_{{\rm A},5, k}+\hat{p}^{(0)}_{{\rm B},5, k-N-V-1}), \nonumber \\
&\hat{\delta}_{{\rm A},5,k} =\frac{{\rm e}^{-r}}{\sqrt{2}}(-\hat{q}^{(0)}_{{\rm A},2, k}-\hat{q}^{(0)}_{{\rm B},2, k-N-V}-\hat{q}^{(0)}_{{\rm A},4, k}-\hat{q}^{(0)}_{{\rm B},4, k-N-V}+\hat{q}^{(0)}_{{\rm A},6, k}+\hat{q}^{(0)}_{{\rm B},6, k-N-V}), \nonumber \\
&\hat{\delta}_{{\rm A},6,k} =\frac{{\rm e}^{-r}}{\sqrt{2}}(\hat{p}^{(0)}_{{\rm A},1, k}+\hat{p}^{(0)}_{{\rm B},1, k-V}+\hat{p}^{(0)}_{{\rm A},3, k}+\hat{p}^{(0)}_{{\rm B},3, k-V}-\hat{p}^{(0)}_{{\rm A},5, k}-\hat{p}^{(0)}_{{\rm B},5, k-V}), \tag{C4}
\end{align}  
and
\begin{align}
&\hat{\delta}_{{\rm B},1,k} =\frac{{\rm e}^{-r}}{\sqrt{2}}(-\hat{q}^{(0)}_{{\rm A},2, k}+\hat{q}^{(0)}_{{\rm B},2, k}-\hat{q}^{(0)}_{{\rm A},4, k}+\hat{q}^{(0)}_{{\rm B},4, k}-\hat{q}^{(0)}_{{\rm A},6, k}+\hat{q}^{(0)}_{{\rm B},6, k}), \nonumber \\
&\hat{\delta}_{{\rm B},2,k}=\frac{{\rm e}^{-r}}{\sqrt{2}}(\hat{p}^{(0)}_{{\rm A},1, k}-\hat{p}^{(0)}_{{\rm B},1, k-1}+\hat{p}^{(0)}_{{\rm A},3, k}-\hat{p}^{(0)}_{{\rm B},3, k-1}+\hat{p}^{(0)}_{{\rm A},5, k}-\hat{p}^{(0)}_{{\rm B},5, k-1}), \nonumber \\
&\hat{\delta}_{{\rm B},3,k} =\frac{{\rm e}^{-r}}{\sqrt{2}}(\hat{q}^{(0)}_{-{\rm A},2, k}+\hat{q}^{(0)}_{{\rm B},2, k-N-1}+\hat{q}^{(0)}_{{\rm A},4, k}-\hat{q}^{(0)}_{{\rm B},4, k-N-1}-\hat{q}^{(0)}_{{\rm A},6, k}+\hat{q}^{(0)}_{{\rm B},6, k-N-1}), \nonumber \\
&\hat{\delta}_{{\rm B},4,k} =\frac{{\rm e}^{-r}}{\sqrt{2}}(\hat{p}^{(0)}_{{\rm A},1, k}-\hat{p}^{(0)}_{{\rm B},1, kv}+\hat{p}^{(0)}_{{\rm A},3, k}-\hat{p}^{(0)}_{{\rm B},3, kv}+\hat{p}^{(0)}_{{\rm A},5, k}-\hat{p}^{(0)}_{{\rm B},5, k-N-V-1}), \nonumber \\
&\hat{\delta}_{{\rm B},5,k} =\frac{{\rm e}^{-r}}{\sqrt{2}}(-\hat{q}^{(0)}_{{\rm A},2, k}+\hat{q}^{(0)}_{{\rm B},2, k-N-V}-\hat{q}^{(0)}_{{\rm A},4, k}+\hat{q}^{(0)}_{{\rm B},4, k-N-V}+\hat{q}^{(0)}_{{\rm A},6, k}-\hat{q}^{(0)}_{{\rm B},6, k-N-V}), \nonumber \\
&\hat{\delta}_{{\rm B},6,k} =\frac{{\rm e}^{-r}}{\sqrt{2}}(\hat{p}^{(0)}_{{\rm A},1, k}-\hat{p}^{(0)}_{{\rm B},1, k-V}+\hat{p}^{(0)}_{{\rm A},3, k}-\hat{p}^{(0)}_{{\rm B},3, k-V}-\hat{p}^{(0)}_{{\rm A},5, k}+\hat{p}^{(0)}_{{\rm B},5, k-V}), \tag{C5}
\end{align}  
respectively. Using $\langle(\hat{q}_{{\rm A(B)},n,k}^{(0)})^2\rangle=\langle(\hat{p}_{{\rm A(B)},n,k}^{(0)})^2\rangle=1/2$, we obtain variances for nullifiers, $\langle \Delta^2 \hat{\delta}_{{\rm A(B)},n,k}\rangle $, as 
\begin{equation}
\langle \Delta^2 \hat{\delta}_{{\rm A(B)},n,k}\rangle =\frac{3}{2}{\rm e}^{-2r}, \tag{C6}
\end{equation}  
and get the inequality 
\begin{equation}
\langle \Delta^2 \hat{\delta}_{{\rm A},1,k}\rangle +\langle \Delta^2  \hat{\delta}_{{\rm A},2,k}\rangle ={\rm e}^{-2r}< \frac{1}{3} \hspace{2pt}, \tag{C7}
\end{equation}
as described in Eq. (33) in the main text.  In the same way as the inequality for qumodes ${\rm A}_{1,k}$ and ${\rm A}_{2,k}$, we can derive the inequality for other qumodes.

\end{widetext}
\end{document}